\pdfoutput=1

\documentclass[reqno]{amsart}
\RequirePackage[l2tabu, orthodox]{nag} 

\usepackage{hyperref}

\usepackage{courier}
\usepackage{color}
\usepackage{amsmath,amsthm,amssymb}
\usepackage{cases} 
\usepackage[cal=boondox,scr=boondoxo]{mathalfa}

\usepackage{graphicx}
\usepackage[figurename=Fig.,labelsep=period]{caption}

\usepackage{textcomp} 
\usepackage{sistyle}
\usepackage{physics}
\usepackage{mathtools}
\mathtoolsset{showonlyrefs=false}

\theoremstyle{definition}
\newtheorem{remark}{Remark}

\usepackage[]{natbib}
\bibpunct[,]{(}{)}{,}{a}{}{,}
\newcommand{\citetapos}[1]{\citeauthor{#1}'s \citeyearpar{#1}}
\newcommand{\citetaposs}[1]{\citeauthor{#1}' \citeyearpar{#1}}

\defcitealias{deMenezes-Barabasi92}{De Menezes and Barab{\'a}si (2004)}

\begin{document}

\title[Scaling of recruitment fluctuations in marine fishes]{
Fluctuation scaling in\\ L{\'e}vy-stable recruitment of marine fishes in\\ randomly varying environments}
\author[H.-S. Niwa]{H.-S. Niwa}
\date{}

\keywords{Taylor's law; random sums; $\alpha$-stable distribution; log-normal distribution; North Atlantic fishes}

\begin{abstract}
This paper studies the scaling properties of recruitment fluctuations in randomly varying environments, for abundant marine species with extreme reproductive behavior. Fisheries stock-recruitment data from the North Atlantic display fluctuation scaling, a proportionality between the standard deviation and the average recruitment among stocks. The proportionality covers over five orders of magnitude in the range studied. A linear-scaling behavior can be a sign of a universal distribution of the normalized data across stocks. In light of this conjecture, it is demonstrated that the L{\'e}vy-stable model offers a better effective description of the recruitment distribution than the log-normal model. Care is devoted to the problem of random sums of random variables. Recruitment is calculated by summing random offspring numbers with infinite variance, where the number of summands (i.e. spawning population size) is also a random process with infinite variance.
\end{abstract}

\maketitle

\section{Introduction}
Marine populations of mass-spawning species are characterized by intermittent, large recruitment events
\citep{Hjort1914}.
Studies on the population dynamics
have been overshadowed by large fluctuations in recruitment
\citep{Smith94}.
Understanding large fluctuations requires theoretical and empirical studies of fluctuation probabilities.
A conventional approach uses the log-normal distribution of recruitment.
While the main body of the recruitment distribution is well described by a log-normal distribution, there are more extreme events than the log-normal model predicts
\citep{Hilborn-Walters92,Niwa06-ALR}.
In this paper, a stable distribution with finite mean and infinite variance
\citep{Khintchine-Levy1936,Levy1937}
is used to describe fluctuations in recruitment,
and applied to time-series data from North Atlantic fish stocks.
I investigate the coupling of recruitment fluctuations with the mean annual recruitment to individual stocks, and find that the coupling scales linearly with the mean recruitment.
Populations with large mean experience proportionally large variability in recruitment,
in contrast to the expectation of decaying per-capita variability.
The linear scaling is here explained by the competition between internal demographic dynamics and changes in external factors (such as climate and commercial exploitation).

Very large interfamilial, or sweepstakes, variation in reproductive success has been documented in abundant marine species
\citep{Hedgecock-Pudovkin2011}.
Population models with skewed offspring-number distributions have recently been proposed as appropriate models to investigate gene genealogies for abundant marine species with type-III (exponential) survivorship curves
\citep{EW06,Sargsyan-Wakeley08}.
Patterns of genetic variation have been studied to search for the imprint of multiple mergers of ancestral lineages
\citep{SBB13,EBBF2015,Arnason-Halldorsdottir2015}.
\citet{Niwa-etal2016} showed that
the multiple-merger coalescent model of the $\mathrm{Beta}(2-\alpha,\alpha)$ type
provides a better fit of Japanese sardine genetic variation
than \citetapos{Kingman82-the-coalescent} classical coalescent model.
The $\mathrm{Beta}(2-\alpha,\alpha)$ coalescent arises from a population model with power-law offspring distribution with tail exponent $1<\alpha<2$
\citep{Schweinsberg2003}.

While the recruiting process has exponential decay in survival probability, the exponential amplification of the number of successfully recruiting offspring (littermates or siblings) in a family compensates the exponentially small probability of their surviving to reproductive maturity.
The combination of these two exponentials leads to power laws in the offspring-number distribution~\citep{Reed-Hughes2002,Newman2005,Niwa-etal2017}.
I consider a population model where the offspring variables are independent copies of random variable $X$ with asymptotic distribution
\begin{equation}\label{eqn:reproductive-variability}
 P(X\geq x)\simeq (x_0/x)^{\alpha},
\end{equation}
which, with $1<\alpha<2$ and $x_0>0$, decays slowly as $x\to\infty$.
The offspring distribution has finite mean (assumed greater than one) and infinite variance.
The recruitment $R$, i.e.
the total number of offspring entering the (potentially reproductive) population, is the sum of $N$ independent random variables $X_1,\ldots,X_N$,
\begin{equation}\label{eqn:recruitment}
 R=\sum_{n=1}^{N}X_n,
\end{equation}
where $X_n$ represents the number of offspring of the $n$-th individual,
and $N$ is the (spawning) population size,
i.e. the number of reproductively-successful individuals in a generation.
This paper considers the limit distribution of the sum $R$, when $N\to\infty$.

If the population size is fixed constant in an environment with constant carrying capacity
(i.e. the offspring generation is constituted by sampling $N$ out of $R$ potential offspring),
the recruitment distribution has mean $\expval{R}=\expval{X}N$, while the squared difference between $R$ and $\expval{R}$
behaves typically as $N^{2/\alpha}$ for large $N$
\citep{Bouchaud-Georges90,Newman2005}.
Thus the typical value of the squared difference is estimated by
\begin{equation}\label{eqn:TL-R-constant-N}
 (R-\expval{R})^2\sim\expval{R}^{2/\alpha},
\end{equation}
whereas the variance $\expval{(R-\expval{R})^2}$ is infinite.
Angled brackets denote mean over all possible realizations.
The value of $\alpha$ in Equation~\eqref{eqn:reproductive-variability} is within the L{\'e}vy-stable domain, suggesting that
the recruitment distribution has similar asymptotic behavior to the offspring distribution.

Stated more precisely, the rescaled sum
\begin{equation*}
 R^{\mathrm{sc}}=\frac{R-\expval{X}N}{\varsigma_R}
\end{equation*}
has a maximally asymmetric $\alpha$-stable distribution $L_{\alpha,1}$ as $N\to\infty$,
where the scale factor $\varsigma_R$ represents the width of the distribution of $R$,
\begin{equation*}
 \varsigma_R=x_0 C_{\alpha} N^{1/\alpha}
\end{equation*}
with
$C_{\alpha}=\left[\Gamma (1-\alpha)\cos(\pi\alpha/2)\right]^{1/\alpha}$.
The probability distribution function $L_{\alpha,\beta}(z)$ is defined by its characteristic function
\begin{equation*}
 \expval{e^{\sqrt{-1}{\,}kz}}=
 \exp\qty[
 -\abs{k}^{\alpha}\qty(1-\sqrt{-1}{\,}\beta\frac{k}{\abs{k}}\tan\qty(\frac{\pi\alpha}{2}))],
\end{equation*}
where
$\alpha$ characterizes the $z\to\infty$ behavior,
\begin{equation*}
 L_{\alpha,\beta}(\pm z)\simeq\frac{1\pm\beta}{2}\alpha z^{-\alpha-1}
\end{equation*}
(double-sign corresponds),
and $\beta\in[-1,1]$ characterizes its degree of asymmetry.
When $\beta=1$, the tail vanishes to the left.
When $\beta=0$, the distribution is symmetric.

Let $R_t$, as in Equation~\eqref{eqn:recruitment}, denote the number of recruits to the population at time $t+1$ in years (or generations),
where $X_n$ is
the number of offspring of the $n$-th individual in year $t$ with the distribution in Equation~\eqref{eqn:reproductive-variability} being identical for all $t$.
I assume sequential independence of events.
The change in recruitment between successive years,
$\Delta R_t=R_{t+1}-R_{t}$,
rescaled by the fluctuation width
$\varsigma_{\Delta R}=x_0 C_{\alpha} (2N)^{1/\alpha}$,
follows a symmetric $\alpha$-stable distribution $L_{\alpha,0}$ as $N\to\infty$.

Equation~\eqref{eqn:TL-R-constant-N} connects the fluctuations to the average recruitment.
This type of scaling relationship is called \citetapos{Taylor61} law or fluctuation scaling \citep{Eisler2008}.
The scaling law states that,
in a set of blocks of observations,
the fluctuations (the sample standard deviation of each block) can be approximately represented by a power-law function of the corresponding average of that block,
\begin{equation*}
 \mbox{fluctuations}\approx\mbox{const.}\times\qty(\mbox{average})^{b}.
\end{equation*}
The scaling exponent $b$ is usually between 1/2 and 1
\citep{Eisler2008}.

Real systems are subject to external forces, such as changing climate and fishing pressure
\citep{Cushing1990,Worm-Myers2004}.
Fish populations in the North Atlantic are buffeted by stochastic harvesting in a random environment, and the population dynamics can be approximately described by a random walk
\citep{Niwa07-ICES,Niwa14}.
\citet{Anderson82} suggested that environmental stochasticity acts to compound the underlying variability created by chance demographic events (births and deaths of individuals).
Here, I study the recruitment fluctuations in randomly varying environments, for abundant marine species with extreme reproductive behavior.
This paper provides, contrary to constant $N$,
a theory for fluctuation scaling of $R$ for a population with a time-variable carrying capacity.
The recruitment is a random sum of independent random variables $X_n$ with infinite variance,
where the number of summands (population size $N$) is also a random process with infinite variance.
Individual reproductive variability (Equation~\ref{eqn:reproductive-variability}) is assumed to be independent of population size.
This paper shows that
fisheries stock-recruitment datasets from the North Atlantic
fall into the `universality' class (or the limiting case) of the exponent $b=1$.
This observation is consistent with \citet{Cobain2019}, which reported the scaling of fluctuations in abundance of populations in the North Sea fish community,
where the exponents are around $b\approx 1$.
The scaling exponent of $b=1$ implies that the normalized observations have an identical distribution across blocks of observations
\citep{Eisler2008}.
In light of this conjecture of a universal distribution, I compare the log-normal versus the L{\'e}vy-stable model for recruitment fluctuations.

The study of random sums of random variables was launched in \citetaposs{Robbins48} paper, where the number $N$ of random variables in a sum is also a random variable.
\citet{Dobrushin55}, \citet{Renyi63}, and \citet{Gnedenko-Fahim69} developed the theory of limiting behavior of random sums.
\citet{Klebanov2012} made certain generalizations of stable distributions of random summands with infinite variances.

This paper contrasts with prior studies that may be related.
\citetalias{deMenezes-Barabasi92} showed that $b=1$ can arise when the external driving force imposes strong fluctuations in $N$,
where both random variables $N$ and $R$ have finite variances.
\citet{Ballantyne-Kerkhoff2007} alternatively suggested that it is possible to obtain $b=1$ for $R$ by sums of identical and completely synchronized $X_n$'s with finite variance
(this phenomenon is called masting in trees).
Some recent results \citep{Cohen2019,Cohen2019-correction} focus on
the sum $R$ of a random number $N$ of correlated random variables $X_n$
with finite mean and variance depending on $N$,
where the mean and variance of $N$ are finite.
For independent random variables $X_n$ distributed as a power law with exponent $\alpha\in (0,1)$, the sample mean and variance of the sum $R$ yield $b=(2-\alpha)/(1-\alpha)>2$ for large fixed $N$
\citep{Bouchaud-Georges90,Newman2005,Brown2017}.
\citet{Fogarty93a} developed alternative expressions for recruitment fluctuations under several stock-recruitment models including Ricker and Beverton-Holt models.

\section{Methods}
\subsection{Recruitment as random sums}
To incorporate externally induced fluctuations, let us allow $N$ to vary independently from one year to the other.
I study the fluctuations of the random sums, allowing for an infinite variance of the number of summands.
The relationship between the fluctuation scaling in the sums (or the value of $b$) and the amplitude of fluctuations in the number of summands is of special interest.

Define $N_t$ ($t=1,\ldots,T$), the size of the population at time $t$, to be a random variable with finite mean $\expval{N}$.
It is assumed that the number of summands is independent of the summands $X_n$.
The average recruitment is then given by
\begin{equation*}
 \expval{R}=\expval{X}\expval{N}.
\end{equation*}
Since the recruitment fluctuations have two sources, it follows that
\begin{equation*}
 \qty(R_t-\expval{R})^2=
 \qty(R_t-\expval{X}N_t)^2
 +\expval{X}^2\qty(N_t-\expval{N})^2
 +2\expval{X}\qty(R_t-\expval{X}N_t)\qty(N_t-\expval{N}),
\end{equation*}
yielding (for large $\sum_{t=1}^T N_t\approx\expval{N}T$)
\begin{equation}\label{eqn:TL-RN}
 \Sigma_R^2\equiv \frac{\sum_{t=1}^T\qty(R_t-\expval{R})^2}{T}
  \approx
  \expval{X}^2 \expval{N}^{2/\alpha} T^{2/\alpha-1}
  +\expval{X}^2\Sigma_N^2
\end{equation}
with
\begin{equation*}
 \Sigma_N^2\equiv \frac{\sum_{t=1}^T\qty(N_t-\expval{N})^2}{T}.
\end{equation*}
When the amplitude of fluctuations $\Sigma_N$ is typically less than $\expval{N}^{1/\alpha}$ and the ratio ${\Sigma_N}/{\expval{N}^{1/\alpha}}$ approaches 0 as $\expval{N}$ gets large,
one recovers Equation~\eqref{eqn:TL-R-constant-N}, i.e.
$\Sigma_R\sim\expval{R}^{1/\alpha}$ for large $\expval{N}$.
In contrast, when $\Sigma_N$ exceeds the threshold, i.e.
the ratio ${\Sigma_N}/{\expval{N}^{1/\alpha}}$ goes to infinity as $\expval{N}\to\infty$,
then the recruitment fluctuations are driven by external forces and one obtains
\begin{equation*}
 \Sigma_R\approx\expval{X}\Sigma_N,
\end{equation*}
which leads to
\begin{equation}\label{eqn:2-linear-functs}
 \Sigma_R/\expval{R}\approx\Sigma_N/\expval{N}.
\end{equation}
Increasing the amplitude of fluctuations $\Sigma_N$ should induce a change from the intrinsic or endogenous to the externally driven behavior.

When there are several stocks (subsystems with respect to the habitat of the species) for which one observes year-class recruitment,
the difference between stocks with smaller and greater mean recruitment comes from the different mean number of spawners.
One is interested in how the sample standard deviation changes, across blocks of observations,
with the value of the mean annual recruitment to individual stocks.
One then calculates the $\expval{R^i}$ and $\Sigma_{R^i}$
(where the superscript $i$ denotes the stock identifier), and compares the couplings across stocks.
If the distribution for the normalized variable, $\nu_t=N_t^i/\expval{N^i}$,
is identical across stocks,
then one has
\begin{equation}\label{eqn:linear-scaling-N}
 \Sigma_{N^i}=\Sigma_{\nu}\expval{N^i}
\end{equation}
with
$\Sigma_{\nu}^2=\overline{[\Sigma_{N^i}^2/\expval{N^i}^2]}$
(average across stocks),
which gives rise to a linear relationship
\begin{equation}\label{eqn:linear-scaling}
 \Sigma_{R^i}\approx\Sigma_{\nu}\expval{R^i}.
\end{equation}
One here sees the scaling exponent $b=1$.
Note that, since $\expval{N^i}$'s are large, the transition between the two scaling regimes is sharp;
see Equations~\eqref{eqn:TL-RN} and~\eqref{eqn:linear-scaling-N}.

When the fluctuations $\Sigma_{N^i}$ of $N^i$'s linearly scale with the mean,
the linear-scaling behavior also arises in $R^i$'s among stocks.
Then, the two linear function, $\Sigma_N$ of $\expval{N}$ and $\Sigma_R$ of $\expval{R}$, nearly overlie each other; see Equation~\eqref{eqn:2-linear-functs}.

Equation~\eqref{eqn:linear-scaling} reflects the central observations in the vast majority rather than rare tail events.
Even though the tail exponents of the offspring distributions are stock-specific,
the $b=1$ behavior can arise among these stocks
when the external driving force imposes strong fluctuations in $N$,
provided that
the normalized variables $N_t^i/\expval{N^i}$ are identically distributed (or almost identically distributed on the central part) across stocks.

\subsection{Asymptotic distribution}
Let $g(\nu_t)$ be the probability distribution function of $\nu_t=N_t/\expval{N}$.
The recruitment $R_t$ is given by
\begin{equation}\label{eqn:transform-R}
 R_t=x_0 C_{\alpha}N_t^{1/\alpha}R_t^{\mathrm{sc}}+\expval{X}\expval{N}\nu_t
\end{equation}
with $R_t^{\mathrm{sc}}$ being distributed as $L_{\alpha,1}$,
where the rescaled variable $R_t^{\mathrm{sc}}$ is independent of $\nu_t$.
When $\Sigma_N/\expval{N}^{1/\alpha}$ tends to 0 as $\expval{N}\to\infty$,
the fluctuations in recruitment are insensitive to external forces,
and the probability distribution function of $R_t$ is described by
\begin{equation}\label{eqn:pdf-R0}
 f(R_t) =
  \qty(x_0 C_{\alpha}\expval{N}^{1/\alpha})^{-1}
  L_{\alpha,1}\qty(\frac{R_t-\expval{X}\expval{N}}{x_0 C_{\alpha}\expval{N}^{1/\alpha}}).
\end{equation}
When $\Sigma_N/\expval{N}^{1/\alpha}$ tends to infinity as $\expval{N}\to\infty$, environmental stochasticity dominates demographic stochasticity.
Then, $R_t/\expval{R}$ is equal in distribution to $N_t/\expval{N}$ for large $\expval{N}$,
and the distribution of $R_t$ is given by
\begin{equation*}
 f(R_t) = \expval{R}^{-1} g\qty(R_t/\expval{R})
\end{equation*}
with $\expval{R}=\expval{X}\expval{N}$.

Assume that the population size $N_t$ is given by
\begin{equation*}
 N_t= \varsigma_N N_t^{\mathrm{sc}} +\expval{N}
\end{equation*}
with scale factor $\varsigma_N$,
where the rescaled variable $N_t^{\mathrm{sc}}$ is distributed according to the $L_{\alpha',\beta}$ with $1<\alpha'<2$ and $-1\leq\beta\leq 1$.
Then, one has
\begin{equation*}
 g(\nu_t) =
  \qty(\varsigma_N/\expval{N})^{-1} L_{\alpha',\beta}\qty(\frac{\nu_t-1}{\varsigma_N/\expval{N}}).
\end{equation*}
When $\varsigma_N/\expval{N}^{1/\alpha}$ tends to infinity as $\expval{N}\to\infty$,
the recruitment distribution has a stable law of index $\alpha'$ and asymmetry parameter $\beta$,
\begin{equation}\label{eqn:pdf-R}
 f(R_t) =
  \qty(\expval{X}\varsigma_N)^{-1}
  L_{\alpha',\beta}\qty(\frac{R_t-\expval{X}\expval{N}}{\expval{X}\varsigma_N})
\end{equation}
for large $\expval{N}$.

The fact that the per-capita variability in population abundance is constant can be a sign of a universal distribution of $N_t^i/\expval{N^i}$ throughout all stocks.
The two distributions of $N_t^i/\expval{N^i}$ and $R_t^i/\expval{R^i}$ then collapse onto a universal curve,
except for large $R_t^i-\expval{R^i}\gtrsim\expval{N^i}$.

\begin{remark}
In the case where year-class strengths are autocorrelated,
if the dependence between distant observations decreases to zero, the successive differences can be represented as
\begin{equation*}
 \Delta R_{t+1} = \lambda\Delta R_t +\varsigma_{\Delta R}z_t
\end{equation*}
with $-1<\lambda<1$,
where $z_t$'s are independent random variables with $L_{\alpha,0}$.
Then, the successive difference reduces to
\begin{equation*}
 \Delta R_t = \sum_{j=1}^{\infty} \varsigma_{\Delta R}\lambda^{j-1} z_{t-j},
\end{equation*}
and one has the characteristic function
\begin{equation*}
 \expval{e^{\sqrt{-1}{\,} k\Delta R}}=
  \exp\qty(-\frac{\varsigma_{\Delta R}^{\alpha}}{1-\lambda^{\alpha}}\abs{k}^{\alpha}).
\end{equation*}
Therefore, the successive differences exhibit symmetric heavy-tail behavior with exponent $\alpha$ and scale $\varsigma_{\Delta R}/(1-\lambda^{\alpha})^{1/\alpha}$.
\end{remark}

\begin{remark}
If $N_t$ has a distribution density which decreases for large $N$ as $N^{-\alpha'-1}$ (with $1<\alpha'<2$),
the largest value $N_{1,T}$ encountered during $T$ independent trials is
$N_{1,T}\sim T^{1/\alpha'}$ as $T\to\infty$.
The amplitude of fluctuations in $N_t$ is
estimated by $\Sigma_N^2\sim T^{2/\alpha'-1}$
\citep{Bouchaud-Georges90,Newman2005}.
Therefore, the sample standard deviation of the recruitment distribution goes as
$\Sigma_R\propto T^{1/\alpha'-1/2}$,
where the endogenous term in Equation~\eqref{eqn:TL-RN} is negligible but the magnitude $\Sigma_N$ is significant.
Although the model is in equilibrium, the recruitment variability increases with observation time $T$,
implying that,
as the number of observations increases, the equilibrium will becomes not more well-defined.
However, for a recruitment series of length $10^2$ (time series in ecology are generally short), one has $T^{1/\alpha'-1/2}<10$ for $1<\alpha'<2$, which is slight for large $N$.
\citep[but see Figure~7 of][]{Halley-Stergiou05}.
\end{remark}

\begin{remark}
Assume that the normalized year-class strengths $R_t^i/\expval{R^i}$ are identically distributed across stocks.
If $R_t^i$ is log-normally distributed, the variance is given by
$\sigma_{R^i}^2=A_{R^i}\expval{R^i}^2$ with
\begin{equation}\label{eqn:lognormal-scaling}
 A_{R^i}=\exp\qty[-2\expval{\ln\qty(R_t^i/\expval{R^i})}]-1.
\end{equation}
Accordingly, a linear relationship between $\sigma_{R^i}$ and $\expval{R^i}$ convincingly emerges
\citep[see also][]{Tippett-Cohen2016}.
\end{remark}

\section{Results}
\subsection{Empirical results}
Time-series data were taken from International Council for the Exploration of the Sea, archived at
\url{http://www.ices.dk/advice/Pages/Latest-Advice.aspx}.
There were 72 fish stocks analyzed throughout the North Atlantic:
19 pelagic stocks (7 species),
50 demersal stocks (16 species),
2 deep-water stocks (2 species)
and one crustacean stock.
The lengths of these time-series varied from 18 to 72 years.
See Appendix \ref{app} for details on the data sources.

\begin{figure}[htb!]
\centering
 \includegraphics[height=.3\textwidth,bb=0 0 360 235]{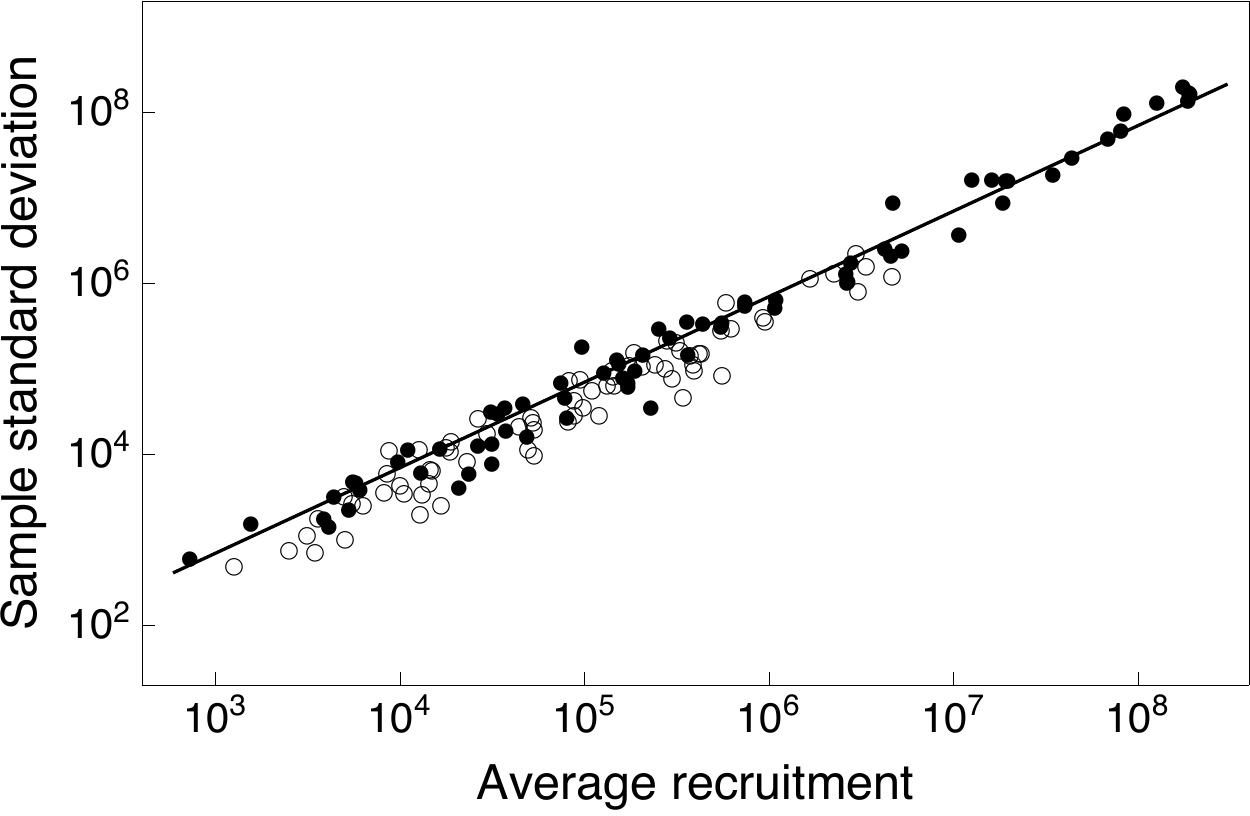}
 \caption{\small
Fluctuation scaling in North Atlantic fish stocks.
The sample standard deviation
$\Sigma_{R^i}$ of each stock $i$ is plotted (on log-log scale in units of $10^3$) in function of the corresponding sample mean number of recruits $\bar{R^i}$ (in thousand fish) of that stock.
The solid line shows the linear function
$y=x\cdot\overline{[\Sigma_{R^i}/\bar{R^i}]}$.
Each open circle corresponds to the average abundance (spawning-stock biomass in tonnes) and sample standard deviation of a single stock.
 }\label{fig:ices2017-TL}
\end{figure}
\subsubsection{Fluctuation scaling}
Figure~\ref{fig:ices2017-TL} shows that the recruitment data (solid circles) across 72 stocks (regardless of species) collapse onto the line
$\Sigma_R\propto\bar{R}$ over five orders of magnitude in $\bar{R}$,
where the overbar denotes sample mean.
Besides, the spawning biomass data (open circles) exhibit constant per-unit-biomass variability, $\Sigma_{N}\propto\bar{N}$.
Notice that $\Sigma_{R}/\bar{R}$ is close to $\Sigma_{N}/\bar{N}$ for each stock,
and these values are approximately constant for all stocks.
The mean across stocks,
\begin{equation*}
 \overline{\qty[\Sigma_{R^i}/\bar{R^i}]}=0.696\quad
  \qty(\mathrm{resp.}{\ }
  \overline{\qty[\Sigma_{N^i}/\bar{N^i}]}=0.472),
\end{equation*}
is closed to the standard deviation of the normalized data (aggregated across stocks),
\begin{equation*}
 \overline{\qty[\qty(R_t^i/\bar{R^i}-1)^2]}^{1/2}=0.746\quad
  \qty(\mathrm{resp.}{\ }
  \overline{\qty[\qty(N_t^i/\bar{N^i}-1)^2]}^{1/2}=0.534),
\end{equation*}
showing that the features captured are not stock-specific.

\subsubsection{Log-normal variation in recruitment}
\begin{figure}[htb!]
 \centering
\includegraphics[height=.3\textwidth,bb=0 0 360 236]{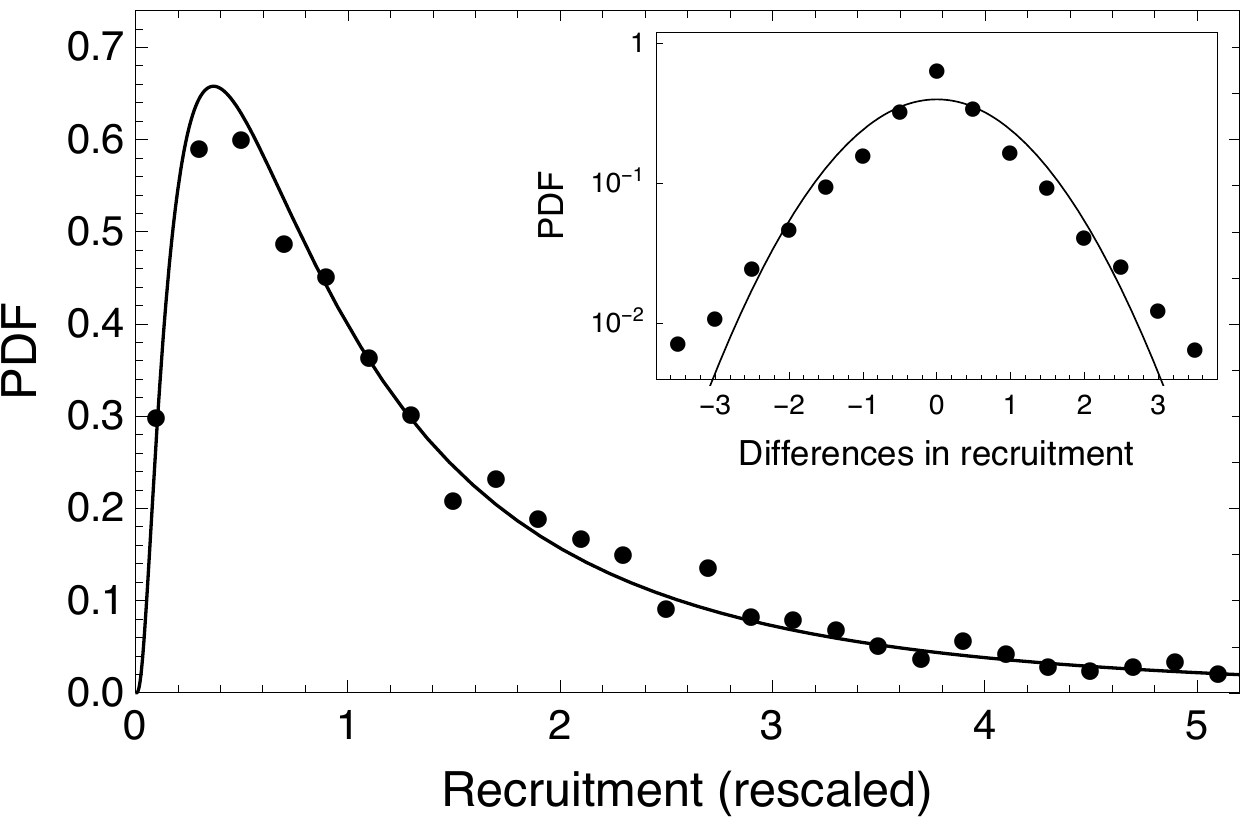}
 \caption{\small
Log-normal distribution.
The recruitment series
(after rescaled with their means and standard deviations of log-transformed data by stocks, and aggregated across stocks)
can be fit by the standard log-normal distribution.
(Inset)
The shape of the distribution of properly rescaled changes in recruitment
is more concentrated than the standard Gaussian distribution,
and the tails are heavier than it,
where, after subtracting the mean of $\Delta R_t^i$, the changes in recruitment are rescaled with the standard deviation by stock.
}\label{fig:ices2017-R-lognormal}
\end{figure}

Figure~\ref{fig:ices2017-R-lognormal} shows that, after rescaled with their means and standard deviations of log-transformed data, the recruitment time-series collapse onto a standard log-normal distribution.

I further study the differences between two consecutive values in (raw) recruitment time-series,
where the original time-series is detrended by subtracting the mean trend.
Based on the Kolmogorov-Smirnov test with data aggregated across 72 stocks,
while the null hypothesis that the (rescaled) log-transformed data are normally distributed cannot be rejected with a $p$-value of $9.54\times 10^{-2}$, the null hypothesis that the successive differences of the (rescaled) recruitment series are normally distributed should be rejected with a $p$-value of $1.14\times 10^{-12}$.
For 36 out of 72 recruitment series that have passed the Ljung-Box test at the 5\% level, which reasonably satisfy the assumption of no autocorrelation, normality of the successive differences is still rejected with
a $p$-value of $1.93\times 10^{-12}$.
While the difference of two independent log-normal random variables should resemble a Gaussian distribution
\citep{Carmona-Durrleman2003},
one sees that rescaled changes in recruitment
have heavier tails than the Gaussian distribution, and they decay more rapidly at the center
(Figure~\ref{fig:ices2017-R-lognormal} inset).

\subsubsection{L{\'evy}-stable distribution for recruitment}
While the log-normal can cover several decades (despite the finiteness of all moments) as long as its standard deviation is sufficiently large,
the Gaussian distribution has rapidly decaying tails.
Therefore, the successive differences of the recruitment series can clarify that
the power-law and log-normal models are distinguishable in their tails.

\begin{figure}[htb!]
 \centering
 \includegraphics[height=.3\textwidth,bb=0 0 360 230]{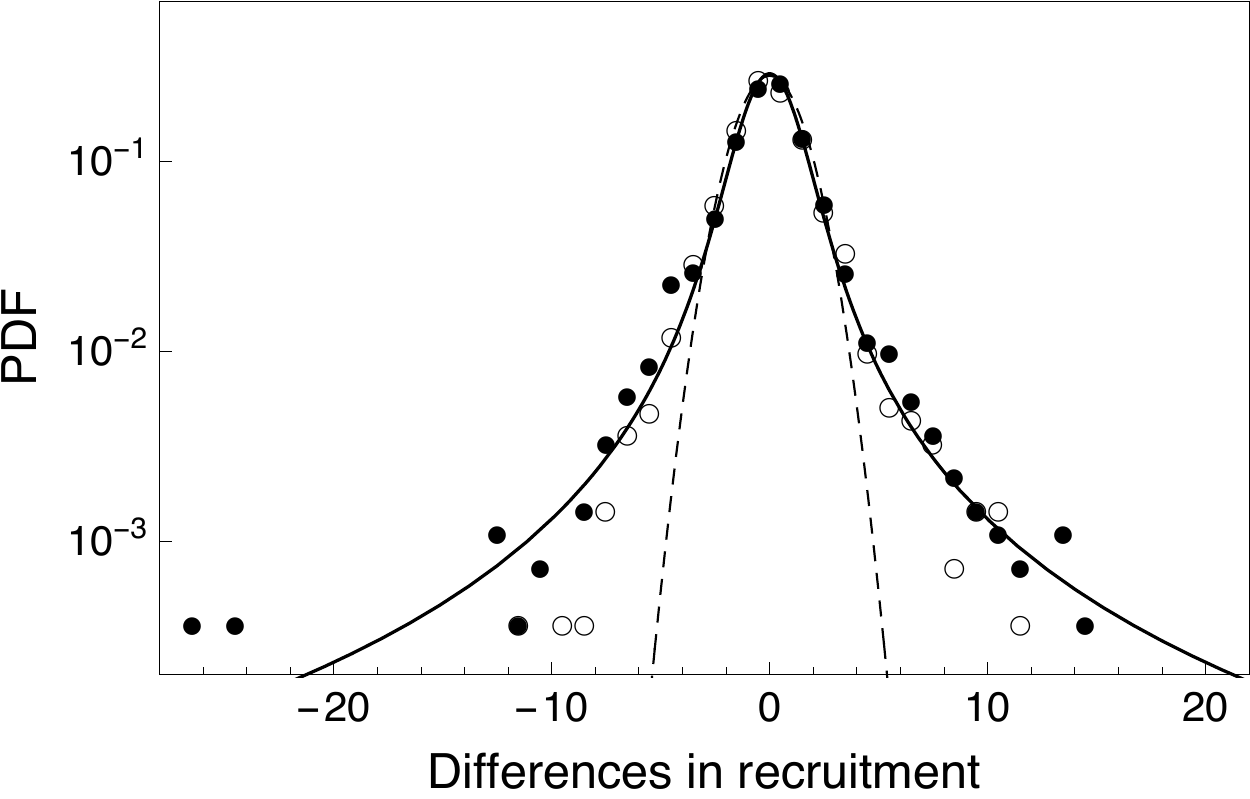}
 \caption{\small
Distribution of successive differences (rescaled with root-median-square values by stocks, and aggregated across stocks).
The differences of successive recruitment series can be characterized by the symmetric $\alpha$-stable distribution $L_{\hat\alpha,0}$ with $\hat\alpha=1.42$ (solid circles and solid line).
The open circles show the distribution of successive differences of the (detrended) biomass series.
The dashed line represents the Gaussian distribution with mean zero and standard deviation $\sqrt{2}$.
}\label{fig:ices2017-R-distribution}
\end{figure}
Figure~\ref{fig:ices2017-TL} suggests that normalized recruitment data are identically distributed across stocks.
I now study the differences between two consecutive values in normalized recruitment series $R_t^i/\bar{R^i}$,
where the time series is detrended by subtracting the mean trend.
Consider the null hypothesis that the successive differences
follow a symmetric $\alpha$-stable distribution.
The successive differences are rescaled with the fluctuation width (i.e. scale factor) by stocks,
where the scale factor is estimated by the root-median-square successive differences of the (detrended) recruitment series
\citep{Takayasu-etal2014}.
Figure~\ref{fig:ices2017-R-distribution} shows that the rescaled distributions with zero mean and unity width, when aggregated across 72 stocks, fit a symmetric $\alpha$-stable distribution with index $\hat\alpha=1.42$ (maximum likelihood estimate MLE);
the null cannot be rejected with a $p$-value of $0.294$ based on the Kolmogorov-Smirnov test.
For the 36 recruitment series showing some evidences of autocorrelation, the null (with the same index $\hat\alpha$) cannot be rejected with a $p$-value of $0.323$;
for the 36 series that have passed the Ljung-Box test, the null cannot be rejected with a $p$-value of $0.314$.
Further, two distributions of $\Delta R_t^i/\bar{R^i}$ and $\Delta N_t^i/\bar{N^i}$, when rescaled with their root-median-square values, collapse onto each other in the central regions.

\subsubsection{Log-normal versus L{\'e}vy}
\begin{figure}[htb!]
 \centering
  \begin{tabular}{ll}
   \small{(a)} & \small{(b)}\\
   \includegraphics[height=.3\textwidth,bb=0 0 360 238]{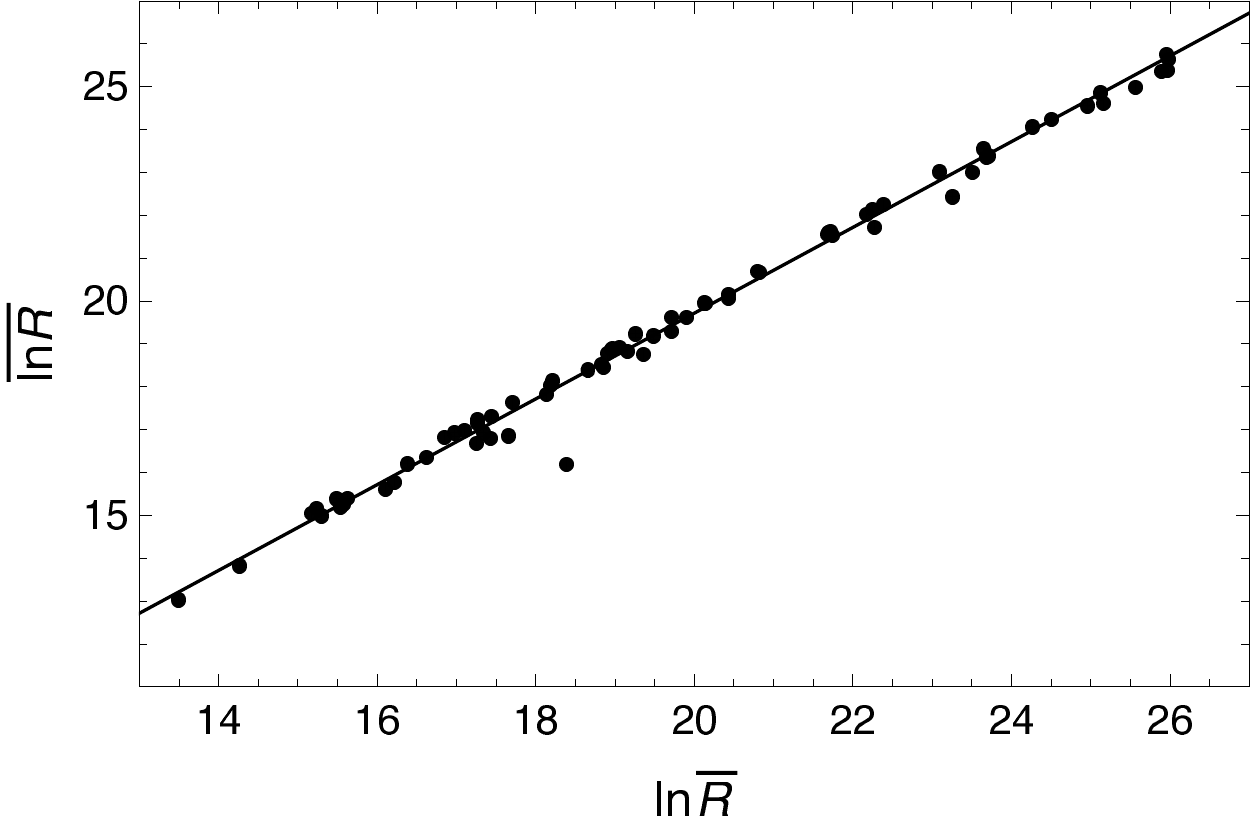}&
       \includegraphics[height=.3\textwidth,bb=0 0 360 235]{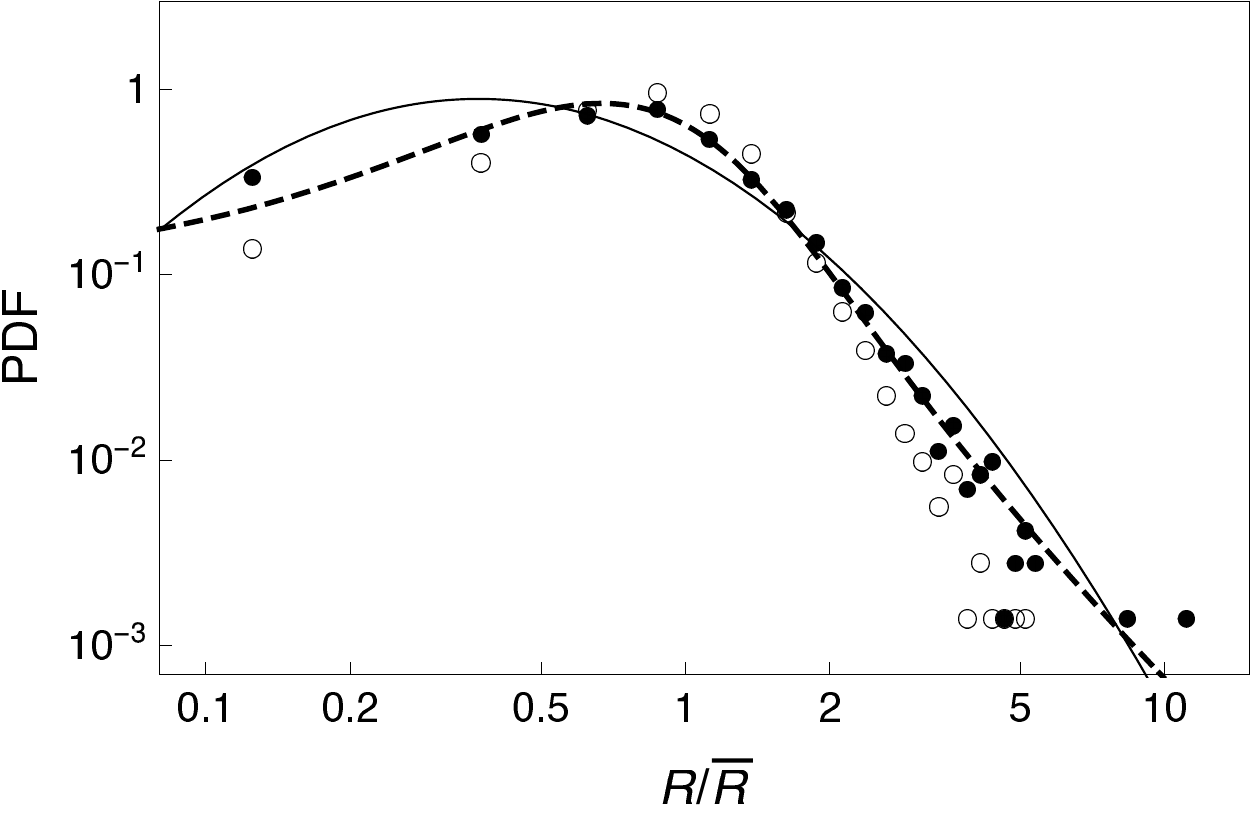}
  \end{tabular}
\caption{\small
Log-normal vs L{\'e}vy.
(a) Arithmetic-geometric means.
Each point corresponds to the log of arithmetic mean $\ln\bar{R^i}$ and log of geometric mean $\overline{\ln R^i}$ of a single stock.
The arithmetic-geometric mean ratio is fairly constant across stocks.
The line shows the linear function
$y=x-\overline{[\ln(A_{R^i}+1)]}/2$
where the mean across stocks
$\overline{[\ln(A_{R^i}+1)]}=0.572$
with $A_{R^i}$ as in Equation~\eqref{eqn:lognormal-scaling}.
(b) Recruitment distribution.
The solid circles show the distribution of normalized recruitment, $R_t^i/\bar{R^i}$, aggregated across stocks.
The solid line is a log-normal distribution, drawn by using the mean and standard deviation of log-transformed data.
 The dashed line represents a maximally asymmetric L{\'e}vy-stable distribution with $\hat\alpha=1.42$ (scale parameter of 0.332 and location parameter of 1.16, MLEs).
The open circles show the distribution of normalized spawning-stock biomass, $N_t^i/\bar{N^i}$, aggregated across stocks.
}\label{fig:ices2017-AG-mean-ratio}
\end{figure}

Figure~\ref{fig:ices2017-AG-mean-ratio}a shows that $A_{R^i}$ in Equation~\eqref{eqn:lognormal-scaling} is independent of $\bar{R^i}$ for North Atlantic fish stocks,
which together with Figure~\ref{fig:ices2017-R-lognormal} might explain the $b=1$ behavior.
Testing the log-normal fit to the distribution of normalized recruitment, $R_t^i/\bar{R^i}$, aggregated across 72 stocks,
I obtain a Kolmogorov-Smirnov $p$-value of $9.03\times 10^{-23}$.
See Figure~\ref{fig:ices2017-AG-mean-ratio}b.

I now verify the fitting of a L{\'e}vy-stable distribution on the normalized recruitment data.
Under the stable law $L_{\hat\alpha,\beta}$ with $\hat\alpha=1.42$,
the degree of asymmetry $\beta$ is estimated at one by maximum likelihood.
The Kolmogorov-Smirnov $p$-value for this maximally asymmetric $\alpha$-stable distribution is $0.179$.

Further, Figure~\ref{fig:ices2017-AG-mean-ratio}b shows that the distribution of the normalized variables $R_t^i/\bar{R^i}$ is close to that for the $N_t^i/\bar{N^i}$ in the central part.
This reflects the fact that the recruitment mimics the environmental (carrying capacity) fluctuations.
Additionally, one observes a difference between the sample mean ($=1$) and the maximum likelihood estimate ($=1.16$) of location.
Note that extreme values of the environmental noise can have a large effect on the sample mean.

\subsection{Numerical results}
Working with the actual infinite-variance stable distributions, I numerically show that the fluctuation scaling holds.
Consider a simple model which consists of two random variables, $N^{\mathrm{sc}}$ and $R^{\mathrm{sc}}$, independently drawn from stable distributions $L_{\alpha',1}$ and $L_{\alpha,1}$, respectively.
The recruitment $R$ is given by Equation~\eqref{eqn:transform-R},
where $\expval{X}=\alpha/(\alpha-1)$ with $x_0=1$ in Equation~\eqref{eqn:reproductive-variability},
and where the population size $N$ is given by
\begin{equation*}
 N_t= \varsigma_N N_t^{\mathrm{sc}} +\mu_N
\end{equation*}
with scale factor $\varsigma_N$ and location parameter $\mu_N$
($=\expval{N}$, mean spawning population sizes).
By setting $\alpha=1.26$ and $\alpha'=1.36$,
and using a population of $\mu_N=10^5$ to $10^{11.5}$,
and three different $\varsigma_N$ values chosen to give
$\varsigma_N=\mu_N^{1/2}$, $\mu_N^{1/\alpha}$ and $0.1\mu_N$,
I sample $10^5$ synthetic data for each  choice of parameter values.
Note that the model does not concern itself with how $N$ individuals in year $t+1$ are sampled (selected for reproduction) out of $R$ extant individuals in year $t$.

\begin{figure}[htb!]
 \centering
 \begin{tabular}{ll}
  \small{(a)} & \small{(b)}\\
  \includegraphics[height=.3\textwidth,bb=0 0 360 231]{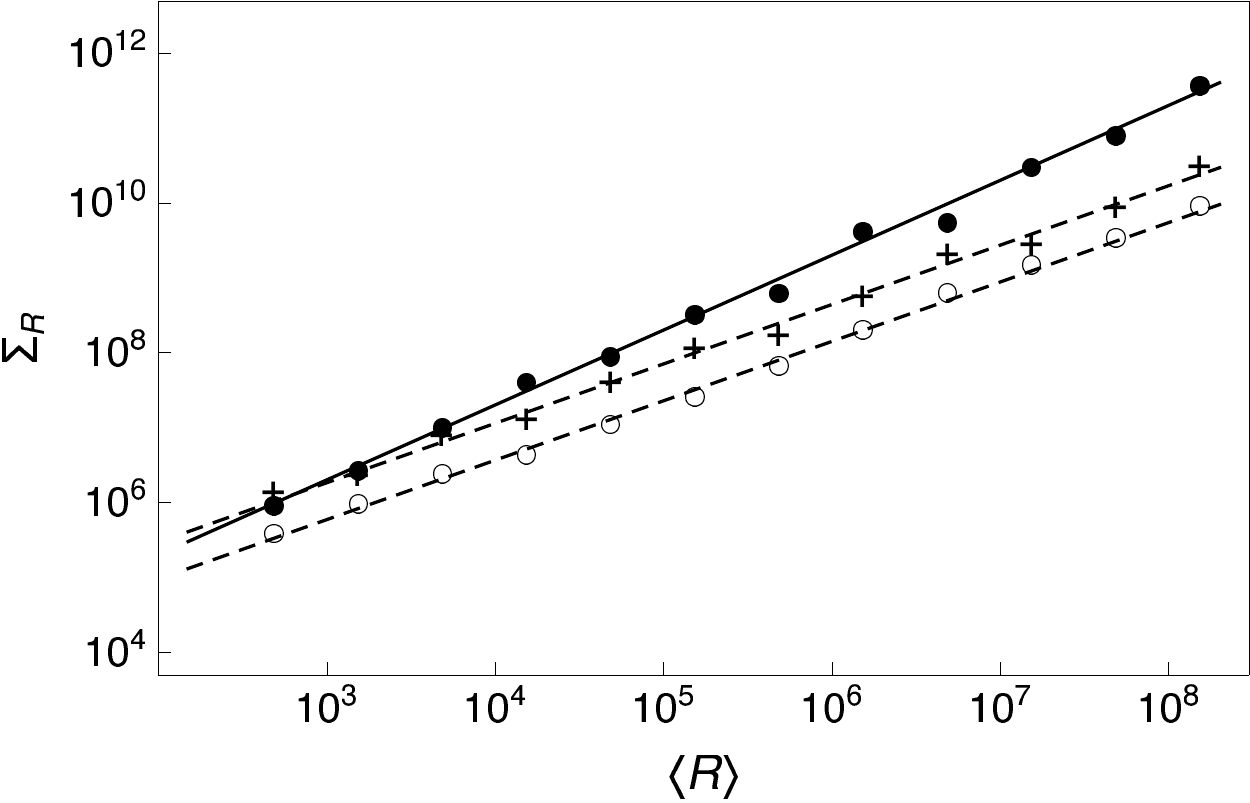}&
      \includegraphics[height=.3\textwidth,bb=0 0 360 230]{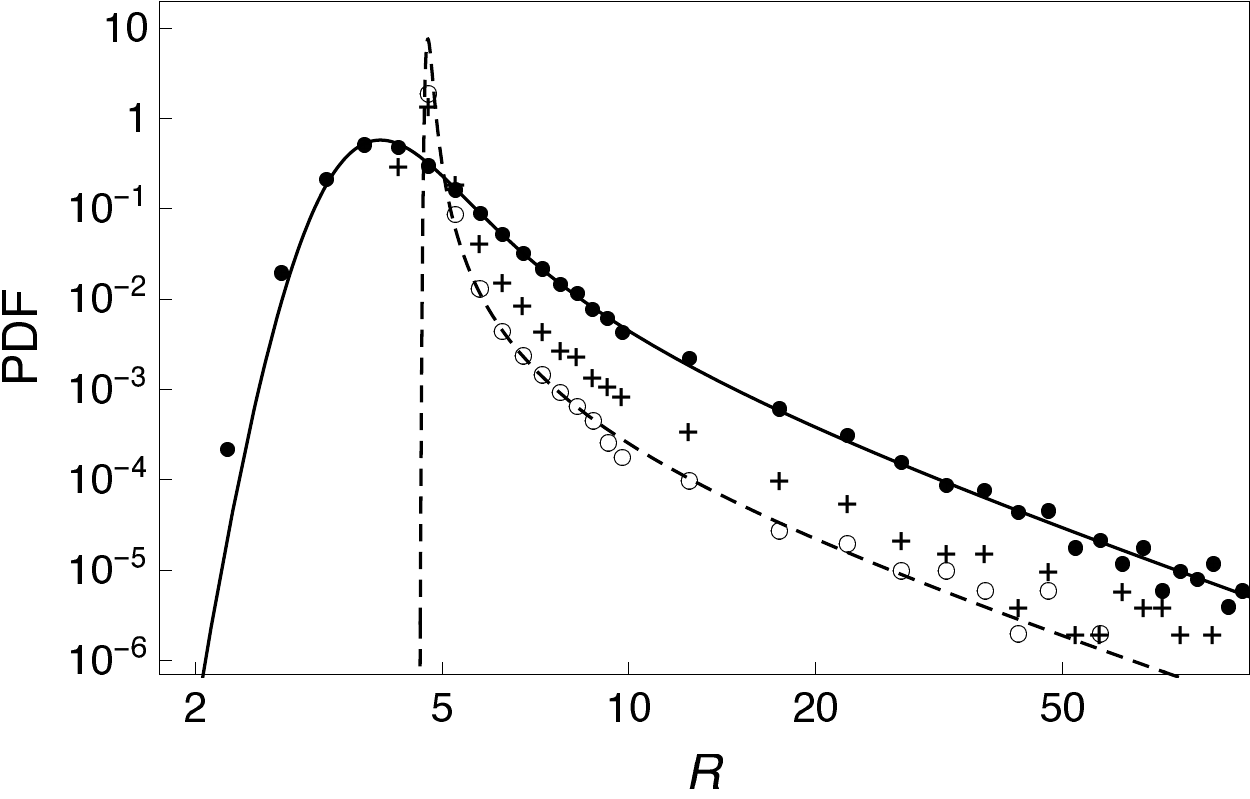}
  \end{tabular}
 \caption{\small
L{\'e}vy-stable model.
(a) Fluctuation scaling (the axes are in units of $10^3$).
Each point corresponds to the coupling between $\expval{R}$ and $\Sigma_R$
under stochastic variation in population sizes,
$\varsigma_N={\expval{N}^{1/2}} {\,\,(\circ)}$,
${\expval{N}^{1/\alpha}} {\,\,(+)}$, and
$0.1{\expval{N}} {\,\,(\bullet)}$.
There is a crossover from $b=1/1.26$
(dashed lines with open circles and plus signs) to $b=1$ (solid line with solid circles),
when the amplitude of the external fluctuations exceeds the threshold.
(b) Recruitment distributions (generated by setting $\expval{N}=10^8$) with $\expval{R}=4.84\expval{N}$.
Points were binned logarithmically ($R$ in unit of $10^8$).
The dashed line represents Equation~\eqref{eqn:pdf-R0} with $\alpha=1.26$.
The solid line represents Equation~\eqref{eqn:pdf-R} with $\alpha'=1.36$, $\beta=1$ and $\varsigma_N=0.1\expval{N}$.
}\label{fig:simulation}
\end{figure}

The case of $\varsigma_N=\mu_N^{1/2}$, where $\Sigma_N/\expval{N}^{1/\alpha}\to 0$ as $\mu_N\to\infty$, shows the feature of a purely demographic process, and one has $\Sigma_R\propto\expval{R}^{1/\alpha}$
(Figure~\ref{fig:simulation}, open circles).
The case of $\varsigma_N=\mu_N^{1/\alpha}$, where typically $\Sigma_N\sim\expval{N}^{1/\alpha}$, shows that
$\Sigma_R$ is still proportional to $\expval{R}^{1/\alpha}$ (Figure~\ref{fig:simulation}, plus signs).
The case of $\varsigma_N=0.1\mu_N$,
where the amplitude of the external fluctuations exceeds the threshold,
shows that $\Sigma_R\propto\expval{R}$ (Figure~\ref{fig:simulation}, solid circles).
After $\Sigma_N$ exceeds the threshold,
the fluctuation $\Sigma_R$ changes the scaling behavior,
and one observes the transition with the crossover from $b=1/\alpha$ to
the universality class of $b=1$.
There are no intermediate $b$ values, since the $b=1/\alpha$ and the $b=1$ scaling do not coexist on the $\Sigma_R$ in Equation~\eqref{eqn:TL-RN} for large $\expval{N}$.

As to the recruitment distribution,
Figure~\ref{fig:simulation}b confirms that as $\expval{N}\to\infty$,
while, when $\Sigma_N/\expval{N}^{1/\alpha}\to 0$, the fluctuations in recruitment are insensitive to external forces,
when $\Sigma_N/\expval{N}^{1/\alpha}\to\infty$, the recruitment mimics the population (carrying capacity) fluctuations.
Equation~\eqref{eqn:pdf-R0} and Equation~\eqref{eqn:pdf-R} agree with the results from the simulations.

\section{Conclusions}
Exploring stylized facts common to fisheries stocks may provide insights into the recruitment variability in exploited populations.
Recruitment time-series across North Atlantic stocks (regardless of species) display fluctuation scaling with universal exponent $b=1$.
The per-recruit variability in recruitment is close to the per-capita variability in population abundance, and both of these are constant for all stocks.
From these facts, it can be deduced that normalized abundances have a universal distribution across stocks,
and consequently normalized year-class strengths have a distribution close to the universal distribution.
I have observed a collapse of these two distributions of the normalized data, aggregated across 72 stocks, onto each other.

Under the conjecture of a universal distribution of the normalized abundances across stocks,
I have explained these stylized facts on recruitment fluctuations in marine fishes
using a simple random-sum model in which the summands are random offspring numbers, and the number of summands (i.e. spawning population size) is also a random process.
When the magnitude of the year-to-year fluctuations in the number of spawners is significant,
the recruitment mimics the environmental (carrying capacity) fluctuations.
Then, one sees the fluctuation scaling in the sums with exponent $b=1$.

The random-sum model sheds new light on models of recruitment distribution.
I have proposed to model the recruitment distribution using a L{\'e}vy-stable law, a natural candidate deriving from the generalized central limit theorem.
The L{\'e}vy-stable model has been shown to provide a better fit to the data than the traditional log-normal model.
The L{\'e}vy-stable distribution has a power-law regime in the tail, which accounts for occurrence of rare large recruitment events.
When aggregated across 72 stocks of North Atlantic fishes, the tail of the recruitment distribution has been found to effectively follow a power law with exponent $\alpha\approx 1.4$.
The fact that $1<\alpha<2$ is consistent with patterns of genetic variation found in marine species
\citep{SBB13,EBBF2015,Arnason-Halldorsdottir2015,Niwa-etal2016}.
Population genetics theory allows the inference of the index $\alpha$ from DNA sequences.

The fluctuations in recruitment of North Atlantic fish stocks are dominated by environmental stochasticity.
It has serious consequences for sustaining the capture fisheries sector.
In the current paradigm of fisheries management,
recruitment variability relative to the mean population abundance $\bar{N}$ is thought to be lower in stocks with more spawning components
\citep{Myers01,Oeberst-etal2009}.
In stark contrast,
the observed proportionality between the mean recruitment $\bar{R}$ and the sample standard deviation $\Sigma_R$ implies that the relative fluctuations $\Sigma_R/\bar{N}$ do not decay with higher mean abundances.
Variation in commercial catches from marine fisheries will have a similar behavior to that of recruitment.
Catch time-series, compiled by the Food and Agriculture Organization of the United Nations, confirmed this expectation
\citep{Niwa06-ecoinf}.

\bibliographystyle{apalike2}
\bibliography{bib-niwa-genetics}

\raggedright
\appendix
\section{Time-series of North Atlantic fish stocks analyzed}\label{app}
Time-series data of 72 fish stocks throughout the North Atlantic were taken from International Council for the Exploration of the Sea.
Data sources are archived at
\url{http://www.ices.dk/advice/Pages/Latest-Advice.aspx}.
The stocks analyzed and the references are as follows:\\[2mm]

\textbf{Pelagic} (19 stocks){\small
\begin{itemize}
 \item[] Capelin (\textit{Mallotus villosus})
         {\footnotesize
         \begin{itemize}
          \item[-] \textsf{cap.27.2a514{\_}Nov} (2017).
                   Iceland and Faroes grounds, East Greenland, Jan Mayen area (November).
                   {doi:10.17895/ices.pub.3616}.
          \item[-] \textsf{cap.27.1-2} (2017).
                   Northeast Arctic (excluding Barents Sea capelin).
                   {doi:10.17895/ices.pub.3272}.
         \end{itemize}
         }
 \item[] Herring (\textit{Clupea harengus})
         {\footnotesize
         \begin{itemize}
          \item[-] \textsf{her.27.1-24a514a} (2017).
                   the Northeast Atlantic and the Arctic Ocean (Norwegian spring-spawning herring).
                   {doi:10.17895/ices.pub.3392}.
          \item[-] \textsf{her.27.3a47d} (2017).
                   North Sea, Skagerrak and Kattegat, and eastern English Channel (autumn spawners).
                   {doi:10.17895/ices.pub.3130}.
          \item[-] \textsf{her.27.28} (2017).
                   Gulf of Riga.
                   {doi:10.17895/ices.pub.3128}.
          \item[-] \textsf{her.27.6a7bc} (2017).
                   West of Scotland, West of Ireland.
                   {doi:10.17895/ices.pub.3061}.
          \item[-] \textsf{her.27.irls} (2017).
                   Irish Sea, Celtic Sea, and southwest of Ireland.
                   {doi:10.17895/ices.pub.3132}.
          \item[-] \textsf{her.27.nirs} (2017).
                   Irish Sea.
                   {doi:10.17895/ices.pub.3133}.
          \item[-] \textsf{her.27.20-24} (2017).
                   Skagerrak, Kattegat, and western Baltic (spring spawners).
                   {doi:10.17895/ices.pub.3126}.
          \item[-] \textsf{her.27.5a} (2017).
                   Iceland grounds (summer-spawning herring).
                   {doi:10.17895/ices.pub.3131}.
          \item[-] \textsf{her.27.25-2932} (2017).
                   central Baltic Sea (excluding the Gulf of Riga).
                   {doi:10.17895/ices.pub.3127}.
          \item[-] \textsf{her.27.3031} (2017).
                   Gulf of Bothnia.
                   {doi:10.17895/ices.pub.3129}.
         \end{itemize}
         }
 \item[] Horse mackerel (\textit{Trachurus trachurus})
         {\footnotesize
         \begin{itemize}
          \item[-] \textsf{hom.27.2a4a5b6a7a-ce-k8} (2017).
                   the Northeast Atlantic.
                   {doi:10.17895/ices.pub.3026}.
          \item[-] \textsf{hom.27.9a} (2017).
                   Atlantic Iberian waters.
                   {doi:10.17895/ices.pub.3062}.
         \end{itemize}
         }
 \item[] Mackerel (\textit{Scomber scombrus})
         {\footnotesize
         \begin{itemize}
          \item[-] \textsf{mac.27.nea} (2017).
                   the Northeast Atlantic and adjacent waters.
                   {doi:10.17895/ices.pub.3023}.
         \end{itemize}
         }
 \item[] Sardine (\textit{Sardina pilchardus})
         {\footnotesize
         \begin{itemize}
          \item[-] \textsf{pil.27.8c9a} (2017).
                   Cantabrian Sea and Atlantic Iberian waters.
                   {doi:10.17895/ices.pub.3307}.
         \end{itemize}
         }
 \item[] Sprat (\textit{Sprattus sprattus})
         {\footnotesize
         \begin{itemize}
          \item[-] \textsf{spr.27.4} (2017).
                   North Sea.
                   {doi:10.17895/ices.pub.3257}.
          \item[-] \textsf{spr.27.22-32} (2017).
                   Baltic Sea.
                   {doi:10.17895/ices.pub.3255}.
         \end{itemize}
         }
 \item[] Blue whiting (\textit{Micromesistius poutassou})
         {\footnotesize
         \begin{itemize}
          \item[-] \textsf{whb.27.1-91214} (2017).
                   Northeast Atlantic and adjacent waters.
                   {doi:10.17895/ices.pub.3030}.
         \end{itemize}
         }
\end{itemize}
}

\textbf{Demersal} (50 stocks){\small
\begin{itemize}
 \item[] White anglerfish (\textit{Lophius piscatorius})
         {\footnotesize
         \begin{itemize}
          \item[-] \textsf{mon.27.8c9a} (2017).
                   Cantabrian Sea and Atlantic Iberian waters.
                   {doi:10.17895/ices.pub.3161}.
         \end{itemize}
         }
 \item[] Cod (\textit{Gadus morhua})
         {\footnotesize
         \begin{itemize}
          \item[-] \textsf{cod.27.6a} (2017).
                   West of Scotland.
                   {doi:10.17895/ices.pub.3100}.
          \item[-] \textsf{cod.27.47d20} (2017).
                   North Sea, eastern English Channel, and Skagerrak.
                   {doi:10.17895/ices.pub.3526}.
          \item[-] \textsf{cod.27.5b1} (2017).
                   Faroe Plateau.
                   {doi:10.17895/ices.pub.3099}.
          \item[-] \textsf{cod.27.22-24} (2017).
                   western Baltic Sea.
                   {doi:10.17895/ices.pub.3095}.
          \item[-] \textsf{cod.27.1-2} (2017).
                   Northeast Arctic.
                   {doi:10.17895/ices.pub.3092}.
          \item[-] \textsf{cod.27.7a} (2017).
                   Irish Sea.
                   {doi:10.17895/ices.pub.3102}.
          \item[-] \textsf{cod.27.7e-k} (2017).
                   eastern English Channel and southern Celtic Seas.
                   {doi:10.17895/ices.pub.3103}.
          \item[-] \textsf{cod.27.5a} (2017).
                   Iceland grounds.
                   {doi:10.17895/ices.pub.3098}.
         \end{itemize}
         }
 \item[] Haddock (\textit{Melanogrammus aeglefinus})
         {\footnotesize
         \begin{itemize}
          \item[-] \textsf{had.27.5a} (2017).
                   Iceland grounds.
                   {doi:10.17895/ices.pub.3119}.
          \item[-] \textsf{had.27.6b} (2017).
                   Rockall.
                   {doi:10.17895/ices.pub.3121}.
          \item[-] \textsf{had.27.5b} (2017).
                   Faroes grounds.
                   {doi:10.17895/ices.pub.3120}.
          \item[-] \textsf{had.27.7b-k} (2017).
                   southern Celtic Seas and English Channel.
                   {doi:10.17895/ices.pub.3123}.
          \item[-] \textsf{had.27.7a} (2017).
                   Irish Sea.
                   {doi:10.17895/ices.pub.3122}.
          \item[-] \textsf{had.27.46a20} (2017).
                   North Sea, West of Scotland, and Skagerrak.
                   {doi:10.17895/ices.pub.3525}.
          \item[-] \textsf{had.27.1-2} (2017).
                   Northeast Arctic.
                   {doi:10.17895/ices.pub.3117}.
         \end{itemize}
         }
 \item[] Hake (\textit{Merluccius merluccius})
         {\footnotesize
         \begin{itemize}
          \item[-] \textsf{hke.27.3a46-8abd} (2017).
                   Greater North Sea, Celtic Seas, and the northern Bay of Biscay (Northern stock).
                   {doi:10.17895/ices.pub.3134}.
          \item[-] \textsf{hke.27.8c9a} (2017).
                   Cantabrian Sea and Atlantic Iberian waters (Southern stock).
                   {doi:10.17895/ices.pub.3135}.
         \end{itemize}
         }
 \item[] Greenland halibut (\textit{Reinhardtius hippoglossoides})
         {\footnotesize
         \begin{itemize}
          \item[-] \textsf{ghl.27.1-2} (2017).
                   Northeast Arctic.
                   {doi:10.17895/ices.pub.3048}.
         \end{itemize}
         }
 \item[] Megrim (\textit{Lepidorhombus whiffiagonis})
         {\footnotesize
         \begin{itemize}
          \item[-] \textsf{meg.27.7b-k8abd} (2017).
                   west and southwest of Ireland, Bay of Biscay.
                   {doi:10.17895/ices.pub.3155}.
          \item[-] \textsf{meg.27.8c9a} (2017).
                   Cantabrian Sea and Atlantic Iberian waters.
                   {doi:10.17895/ices.pub.3156}.
         \end{itemize}
         }
 \item[] Four-spot megrim (\textit{Lepidorhombus boscii})
         {\footnotesize
         \begin{itemize}
          \item[-] \textsf{ldb.27.8c9a} (2017).
                   southern Bay of Biscay and Atlantic Iberian waters East.
                   {doi:10.17895/ices.pub.3152}.
         \end{itemize}
         }
 \item[] Plaice (\textit{Pleuronectes platessa})
         {\footnotesize
         \begin{itemize}
          \item[-] \textsf{ple.27.21-23} (2017).
                   Kattegat, Belt Seas, and the Sound.
                   {doi:10.17895/ices.pub.3049}.
          \item[-] \textsf{ple.27.420} (2017).
                   North Sea and Skagerrak.
                   {doi:10.17895/ices.pub.3529}.
          \item[-] \textsf{ple.27.7d} (2017).
                   eastern English Channel.
                   {doi:10.17895/ices.pub.3200}.
          \item[-] \textsf{ple.27.7a} (2017).
                   Irish Sea.
                   {doi:10.17895/ices.pub.3198}.
         \end{itemize}
         }
 \item[] Norway pout (\textit{Trisopterus esmarkii})
         {\footnotesize
         \begin{itemize}
          \item[-] \textsf{nop.27.3a4} (2017).
                   North Sea, Skagerrak, and Kattegat.
                   {doi:10.17895/ices.pub.4241}.
         \end{itemize}
         }
 \item[] Beaked redfish (\textit{Sebastes mentella})
         {\footnotesize
         \begin{itemize}
          \item[-] \textsf{reb.27.1-2} (2017).
                   Northeast Arctic.
                   {doi:10.17895/ices.pub.3212}.
         \end{itemize}
         }
 \item[] Golden redfish (\textit{Sebastes norvegicus})
         {\footnotesize
         \begin{itemize}
          \item[-] \textsf{reg.27.561214} (2017).
                   Iceland and Faroes grounds, West of Scotland, North of Azores, and East of Greenland.
                   {doi:10.17895/ices.pub.3215}.
          \item[-] \textsf{reg.27.1-2} (2018).
                   Northeast Arctic.
                   {doi:10.17895/ices.pub.4408}.
         \end{itemize}
         }
 \item[] Saithe (\textit{Pollachius virens})
         {\footnotesize
         \begin{itemize}
          \item[-] \textsf{pok.27.5a} (2017).
                   Iceland grounds.
                   {doi:10.17895/ices.pub.3207}.
          \item[-] \textsf{pok.27.1-2} (2017).
                   Northeast Arctic.
                   {doi:10.17895/ices.pub.3205}.
          \item[-] \textsf{pok.27.3a46} (2017).
                   North Sea, Rockall and West of Scotland, Skagerrak and Kattegat.
                   {doi:10.17895/ices.pub.3206}.
          \item[-] \textsf{pok.27.5b} (2017).
                   Faroes grounds.
                   {doi:10.17895/ices.pub.3208}.
         \end{itemize}
         }
 \item[] Sandeel (\textit{Ammodytes} spp.)
         {\footnotesize
         \begin{itemize}
          \item[-] \textsf{san.sa.4} (2017).
                   northern and central North Sea.
                   {doi:10.17895/ices.pub.2680}.
          \item[-] \textsf{san.sa.3r} (2017).
                   northern and central North Sea, Skagerrak.
                   {doi:10.17895/ices.pub.3226}.
          \item[-] \textsf{san.sa.2r} (2017).
                   central and southern North Sea.
                   {doi:10.17895/ices.pub.3225}.
          \item[-] \textsf{san.sa.1r} (2017).
                   central and southern North Sea, Dogger Bank.
                   {doi:10.17895/ices.pub.3034}.
         \end{itemize}
         }
 \item[] Seabass (\textit{Dicentrarchus labrax})
         {\footnotesize
         \begin{itemize}
          \item[-] \textsf{bss.27.4bc7ad-h} (2017).
                   central and southern North Sea, Irish Sea, English Channel, Bristol Channel, and Celtic Sea.
                   {doi:10.17895/ices.pub.3334}.
         \end{itemize}
         }
 \item[] Sole (\textit{Solea solea})
         {\footnotesize
         \begin{itemize}
          \item[-] \textsf{sol.27.4} (2017).
                   North Sea.
                   {doi:10.17895/ices.pub.3528}.
          \item[-] \textsf{sol.27.20-24} (2017).
                   Skagerrak and Kattegat, western Baltic Sea.
                   {doi:10.17895/ices.pub.3229}.
          \item[-] \textsf{sol.27.7fg} (2017).
                   Bristol Channel and Celtic Sea.
                   {doi:10.17895/ices.pub.3234}.
          \item[-] \textsf{sol.27.7e} (2017).
                   western English Channel.
                   {doi:10.17895/ices.pub.3233}.
          \item[-] \textsf{sol.27.8ab} (2017).
                   northern and central Bay of Biscay
                   {doi:10.17895/ices.pub.3236}.
          \item[-] \textsf{sol.27.7d} (2017).
                   eastern English Channel.
                   {doi:10.17895/ices.pub.3232}.
          \item[-] \textsf{sol.27.7a} (2018).
                   Irish Sea.
                   {doi:10.17895/ices.pub.4482}.
         \end{itemize}
         }
 \item[] Whiting (\textit{Merlangius merlangus})
         {\footnotesize
         \begin{itemize}
          \item[-] \textsf{whg.27.47d} (2017).
                   North Sea and eastern English Channel.
                   {doi:10.17895/ices.pub.3530}.
          \item[-] \textsf{whg.27.7a} (2017). 
                   Irish Sea.
                   {doi:10.17895/ices.pub.3268}.
          \item[-] \textsf{whg.27.7b-ce-k} (2017).
                   southern Celtic Seas and eastern English Channel.
                   {doi:10.17895/ices.pub.3269}.
          \item[-] \textsf{whg.27.6a} (2018).
                   West of Scotland.
                   {doi:10.17895/ices.pub.4484}.
         \end{itemize}
         }
\end{itemize}
}

\textbf{Deep-water} (2 stocks){\small
\begin{itemize}
 \item[] Ling (\textit{Molva molva})
         {\footnotesize
         \begin{itemize}
          \item[-] \textsf{lin.27.5a} (2017).
                   Iceland grounds.
                   {doi:10.17895/ices.pub.3139}.
         \end{itemize}
         }
 \item[] Tusk (\textit{Brosme brosme})
         {\footnotesize
         \begin{itemize}
          \item[-] \textsf{usk.27.5a14} (2017).
                   East Greenland and Iceland grounds.
                   {doi:10.17895/ices.pub.3266}.
         \end{itemize}
         }
\end{itemize}
}

\textbf{Crustacean} (1 stock){\small
\begin{itemize}
 \item[] Northern shrimp (\textit{Pandalus borealis})
         {\footnotesize
         \begin{itemize}
          \item[-] \textsf{pra.27.4a20-Oct} (2017).
                   northern North Sea, in the Norwegian Deep and Skagerrak.
                   {doi:10.17895/ices.pub.3357}.
         \end{itemize}
         }
\end{itemize}
}

\end{document}